\documentclass[showpacs]{revtex4}
\usepackage[dvips]{color}
\usepackage{graphicx}
\usepackage{amsmath}

\def\lsim{\mathrel{\raise2pt\hbox to 8pt{\raise -5pt\hbox{$\sim$}\hss{$<$}}}}
\def\bmath#1{\mbox{\boldmath $#1$}}

\begin{document}
\title{Pseudospin symmetry as a relativistic dynamical symmetry in the
nucleus}
\author{P. Alberto}
\affiliation{Departamento de F\'\i sica and Centro de F\'\i sica Computacional,
Universidade de Coimbra, P-3004-516 Coimbra Portugal}
\author{M. Malheiro}
\affiliation{Instituto de F\'\i sica, Universidade Federal Fluminense, \\
24210-340 Niter\'oi, Brazil}
\author{A. Delfino}
\affiliation{Instituto de F\'\i sica, Universidade Federal Fluminense, \\
24210-340 Niter\'oi, Brazil}  
\author{M. Fiolhais}
\affiliation{Departamento de F\'\i sica and Centro de F\'\i sica Computacional,
Universidade de Coimbra, P-3004-516 Coimbra Portugal}
\author{M. Chiapparini}
\affiliation{Departamento de F\'\i sica Te\'orica, Universidade do Estado
 do Rio de Janeiro,\\
20550-900 Rio de Janeiro, Brazil}
\pacs{21.10.-k, 21.10.Hw, 21.60.Cs}
\date{\today}

\begin{abstract} 
\noindent
Pseudospin symmetry in nuclei is investigated by solving the
Dirac equation with Woods-Saxon scalar and vector radial potentials, 
and studying the 
correlation of the energy splittings of pseudospin partners
with the nuclear potential parameters. 
The pseudospin interaction is related to a pseudospin-orbit term
that arises in a Schroedinger-like equation for the lower component
of the Dirac spinor. We show that the contribution from this term to the 
energy splittings of pseudospin partners is large. The near pseudospin degeneracy
results from a significant cancelation among the different terms in that equation, 
manifesting the dynamical character of this symmetry in the nucleus.
 We analyze the isospin
dependence of the pseudospin symmetry and find that its dynamical
character is behind the different pseudospin splittings
observed in neutron and proton spectra of nuclei.
\end{abstract}
\maketitle

\section{Introduction}

The idea of pseudospin was introduced to explain the 
quasi-degeneracy in some nuclei
between single-nucleon states with quantum numbers
($n$, $\ell$, $j= \ell + 1/2$) and
($n-1$, $\ell+2$, $j= \ell$ + 3/2)
where $n$, $\ell$, and
$j$ are the radial, the orbital, and the total angular momentum
quantum numbers, respectively. These levels have the
same ``pseudo'' orbital angular momentum quantum number,
$\tilde{\ell} = \ell + 1$, and ``pseudo'' spin quantum number,
$\tilde s = 1/2$.
For example, for $[n s_{1/2},(n-1) d_{3/2}]$ one has $\tilde{\ell}= 1$,
for $[n p_{3/2},(n-1) f_{5/2}]$ one has $\tilde{\ell}= 2$, etc.
Pseudospin symmetry is exact when
doublets with $j = \tilde{\ell}\ \pm \tilde s$ are degenerate. 

Since the proposal of the pseudospin concept, about 30 years ago
\cite{kth,aa}, the underlying mechanism
of pseudospin symmetry in nuclei has become
a topic of intense theoretical research. In a recent letter
\cite{pmmdm_prl}, we have shown
the dynamical character of this symmetry in nuclei, arising from
balance effects among the central nuclear potentials parameters. 
We have also concluded that this dynamical behavior of the
symmetry is the main explanation for the isospin asymmetry of the
nuclear pseudospin orbit interaction observed in nature. In this paper we
elaborate on the main topics of that letter,
extending our analysis for different nuclei.

It is worth reviewing the main results of several
papers \cite{bahri,mosh,mosk,gino,arima,meng0}
in which the possible underlying mechanisms to generate pseudospin
symmetry have been discussed.

A helicity unitary transformation of a non-relativistic
single-particle Hamiltonian was considered by
Blokhin {\em et al.} \cite {bahri}. They showed that 
the transformed radial wave functions behave as
\begin{equation}
\tilde{R}_{nj\tilde{\ell}} \sim \left\{
\begin{array}{ll}
 r^{\tilde{\ell}}, & r \rightarrow 0 \, , \\
 r^{-(\ell+3)},    & r \rightarrow \infty \, .
\end{array}
\right.
\label {blokhin}
\end{equation}
Moreover, they claimed this behavior to be universal and conjectured that, at
small distances, the helicity transformed mean field of a heavy nucleus
should not be much different from the untransformed one, since the
behavior of $\tilde{R}$ remains similar to $R$ near the
origin.  At larger distances, however, as suggested by the above
asymptotic behavior of the transformed wave function, the helicity
transformed mean field would acquire a more diffuse surface.
At the same time, the helicity transformed fields would
get strongly nonlocal at
the surface.  The application of the helicity operator to the
non-relativistic single-particle wave function maps the normal state
$(\ell,s)$ onto the ``pseudo" state $(\tilde\ell, \tilde s )$, while
keeping all other global symmetries \cite{bahri}. 

A similar kind of unitary
transformation was also considered earlier to discuss the pseudospin
symmetry in the non-relativistic harmonic oscillator \cite{mosh,mosk}.
It was shown that, if the ratio of the coefficients of spin-orbit and 
orbit-orbit terms in the single-particle spherical Nilsson Hamiltonian 
has a particular value, pseudospin symmetry will be exact. Bahri {\em et al.} 
\cite{mosk} showed that this value was approached by relativistic mean field (RMF)
estimates for that ratio.  

More recently, the subject was revived when Ginocchio \cite {gino}
recognized, for the first time, the relativistic character of the symmetry.
He noted that the pseudo-orbital angular momentum is just the orbital 
angular momentum of the lower component of the Dirac spinor. He also showed that
pseudospin symmetry is an exact symmetry for the Dirac Hamiltonian with an 
attractive scalar potential, $S$, and a repulsive vector potential $V$, 
when these potentials are equal in magnitude: $S+V=0$. 
It turns out that this symmetry is nothing but a SU(2) symmetry of the 
Dirac Hamiltonian \cite{levi,smith,bell}.

In RMF models, often referred to as Quantum
Hadrodynamics \cite{wal}, the nuclear saturation mechanism is explained
by the cancelation between a large scalar and a large
vector fields \cite {wal,furn0,zim,delf,delf1}.  Typical values for
these fields in heavy nuclei are of the order of a few hundred MeV
(with opposite signs), their sum providing a binding potential of about
$-60$~MeV at the center.  Therefore, it was natural for Ginocchio to
regard the quasi-degenerate pseudospin doublets in nuclei as arising
from the near equality in magnitude of the attractive scalar, $S$, and
the repulsive vector, $V$, relativistic mean fields, $S \sim - V$, in
which the nucleons move.

After the work of Ginocchio, Meng {\em et al.} \cite{arima,meng0} showed that
pseudospin symmetry is exact when $d \Sigma/dr=0$, where $\Sigma=S+V$.
They also related the onset of the pseudospin symmetry as
a competition between the centrifugal barrier and the
pseudospin-orbit potential.  In short, they claimed that the following
inequality assures pseudospin symmetry:
\begin{equation}
\frac{\kappa}{r} \frac{d \Sigma(r)}{ d r} \ll [E-\Sigma(r)] \frac{\kappa (\kappa-1)}
{r^2}\, ,
\label{arim}
\end{equation}
where the quantum number $\kappa$ obeys
$\kappa(\kappa-1)\,\,=\,\,\ell(\ell+1)$ and $E$ is the binding energy.  
The above inequality is, of course,
satisfied in the particular cases $\Sigma=0$ and $d \Sigma/dr\,=\,0$.
One may note that the latter condition implies the former if
$S,V\rightarrow 0$ at large distances, which is the case of the nucleus.
However, this criterium presents several practical problems that are not
easily overcome \cite{marcos}, namely it cannot be applied
in the range of values of 
$r$ for which $E\sim \Sigma(r)$. This is related to the fact that the 
pseudospin-orbit term has the denominator $E-\Sigma(r)$
and thus becomes infinite when $E=\Sigma(r)$ \cite{arima,meng0,marcos}.

In other works \cite{ring,gino2,gino3,arima1,arima2} the pseudospin
symmetry has also been discussed in relation with the arguments and
conjectures of refs.  \cite{gino,arima,meng0}. Recently, a test of nuclear 
wave functions for pseudospin symmetry was done in \cite{gino5}.
 In a more recent work, the structure of radial nodes occuring in pseudospin levels and a classification for the intruder levels (states which do not have a pseudospin
partner in the limit of pseudospin symmetry) was explained as a direct effect of the 
behavior of nodes of Dirac bound states \cite{levi1}, giving support for the
relativistic interpretation of nuclear pseudospin symmetry. However, despite these
reports on the quasi-degeneracy of pseudospin doublets, the nature of pseudospin
breaking in nuclei is not fully understood.

In this paper we address this question by studying the splittings
of the neutron single-particle energy levels for pseudospin partners
obtained by solving the Dirac equation with vector and scalar potentials of 
Woods-Saxon type.
From the behavior of these splittings with Woods-Saxon parameters, we
are able to show that pseudospin symmetry has a dynamical character.
As we have already pointed out, the present paper is a natural extension
of \cite{pmmdm_prl}, providing more technical details and complementary
physical information. In addition, we calculate the
expectation value of the pseudospin-orbit coupling and find that it correlates
well with the observed pseudospin splittings.

The paper is organized as follows:
in Section II we review the formalism of the Dirac equation with scalar
and vector potentials, its relation to pseudospin symmetry and
identify the pseudospin-orbit term. The 
neutron energy levels and respective pseudospin splittings
obtained in a relativistic calculation with Woods-Saxon
mean fields are presented in Section III. We consider the cases
of $^{208}$Pb, $^{40}$Ca and $^{48}$Ca and draw the systematics for
the variation of pseudospin splittings with the Woods-Saxon potential parameters. 
We also study the contribution to the energy splitting of pseudospin partners by each term in the Schroedinger-like second order differential
equation for the lower component of the Dirac spinor, in particular
the contribution of the pseudospin-orbit term. 
Using this systematics, in Section IV we explain the differences of 
those splittings for neutrons and protons and why this does not happen 
with spin-orbit splittings. In Section V we give a brief summary of our main conclusions.

\section{Dirac equation and pseudospin symmetry}

The Hamiltonian of a Dirac particle of
mass $m$ in an external scalar, $S$, and vector,
$V$, potentials is given by
\begin{eqnarray}
H = \bmath{\alpha}\cdot\bmath p
+ \beta (m  + S) + V  ~,
\label {dirac}
\end{eqnarray}
where ${\bmath\alpha}$ and $\beta $ are the Dirac
matrices
\begin{equation}
\beta=
\left(\begin{array}{cc}I&0\cr 0& -I\\\end{array}\right)\ ,
\qquad\bmath\alpha=
\left(\begin{array}{cc}0&\bmath\sigma\\ \bmath\sigma&0\\ \end{array}\right)\ ,
\end{equation}
and $\bmath{\sigma}$ are the Pauli matrices.

The Hamiltonian (\ref{dirac}) is invariant under SU(2)
transformations for two cases:  $S$ = $V$ and $S = - V$
\cite{smith,bell}.  The first case was recently invoked to explain the
spectrum of some mesons, exhibiting small spin-orbital splittings \cite
{gino4}.  The second situation would lead to exact pseudospin symmetry in
nuclei, with $\Sigma=0$. However, since $\Sigma$ plays the role of a binding potential,
this value for $\Sigma$ would imply that there would be no bound nucleon states \cite{gino}.
In relativistic mean field models the attractive scalar and the
repulsive vector parts of the nucleon potential are similar in magnitude
but they are not equal, and so that limiting case is never reached.

 We define 
$\Delta=V-S$ and denote the upper and lower components of the
Dirac spinor by $\Psi_{\pm}=\frac{1 \pm \beta}{2}\Psi$.
The Dirac equation $H\Psi=\epsilon\Psi$ with $H$ given by (\ref{dirac})
and energy $\epsilon=E+m$ gives rise to two coupled first-order differential equations
for the upper and lower components:
\begin{subequations}
\label{dirac_first}
\begin{eqnarray}
\label{dirac_first1}
\bmath{\alpha}\cdot\bmath p\,\Psi_-&=&(E-\Sigma)\Psi_+\\
\label{dirac_first2}
\bmath{\alpha}\cdot\bmath p\,\Psi_+&=&(E+2m-\Delta)\Psi_-\ .
\end{eqnarray}
\end{subequations}
If one multiplies these equations by $\bmath{\alpha}
\cdot\bmath p$ one gets
\begin{eqnarray}
p^2\Psi_{-}&=&-[\bmath{\alpha}\cdot\bmath p\,\Sigma]\Psi_{+}+(E-\Sigma)
\bmath{\alpha}\cdot\bmath p\,\Psi_{+}
\label{ps}
\\
p^2\Psi_{+}&=&-[\bmath{\alpha}\cdot\bmath p\,\Delta]\Psi_{-}+(E+2m-\Delta)
\bmath{\alpha}\cdot\bmath p\,\Psi_{-}\ ,
\label{ps1}
\end{eqnarray}
where the square brackets mean that the operator $\bmath{\alpha}
\cdot\bmath p$ only acts on the potential in front of it. Using 
Eqs.~(\ref{dirac_first}) one obtains
\begin{eqnarray}
p^2\Psi_{-}&=&-[\bmath{\alpha}\cdot\bmath p\,\Sigma]\Psi_{+}+
(E+2m-\Delta)(E-\Sigma)\Psi_-
\label{pss}
\\
p^2\Psi_{+}&=&-[\bmath{\alpha}\cdot\bmath p\,\Delta]\Psi_{-}+(E+2m-\Delta)
(E-\Sigma)\Psi_+\ .
\label{pss1}
\end{eqnarray}

Assuming that the potentials $\Sigma$ and 
$\Delta$ are radial functions, and using again Eqs.~(\ref{dirac_first}),
we obtain the following two decoupled second-order 
differential equations for $\Psi_{\pm}$:
\begin{eqnarray}
p^2\Psi_{-}&=&-\frac{\Sigma'}{E-\Sigma}\bigg(-\frac{\partial\hfill}{\partial r}+
\frac{1}{r} \bmath{\sigma}\cdot\bmath{L}\bigg)\Psi_{-}
+(E+2m-\Delta)(E-\Sigma)\Psi_{-}
\label{psiminus}
\\
p^2\Psi_{+}&=&-\frac{\Delta'}{E+2m-\Delta}\bigg(-\frac{\partial\hfill}{\partial r}+
\frac{1}{r} \bmath{\sigma}\cdot\bmath{L}\bigg)\Psi_{+}
+(E+2m-\Delta)(E-\Sigma)\Psi_{+} \ ,
\end{eqnarray}
where the primes denote derivatives with
respect to $r$. The spinors
$\Psi_\pm$ can be factorized into radial and angular parts:
$\Psi_+={\rm i}\,G_i(r)\,\Phi^+_i(\theta,\phi)$ and
$\Psi_-=-F_i(r)\,\Phi^-_i(\theta,\phi)$,
where $i$ denotes the quantum numbers of the single particle state.
In terms of the radial functions $G_i$ and $F_i$ for upper and lower
components, these equations become
\begin{eqnarray}
\nabla^2 F_i+\frac{\Sigma'}{E-\Sigma}\bigg(F'_i+
\frac{1-\kappa_i}{r}F_i\bigg)
+(E+2m-\Delta)(E-\Sigma)F_i=0
\label {lower}
\\
\nabla^2 G_i+\frac{\Delta'}{E+2m-\Delta}\bigg(G'_i+
\frac{1+\kappa_i}{r}G_i\bigg)
+(E+2m-\Delta)(E-\Sigma)G_i=0 \, ,
\label{upper}
\end{eqnarray}
where  the property
$\bmath{\sigma}\cdot\bmath{L}\Phi^\pm_i=-(1\pm\kappa_i)\Phi^\pm_i$ was used.
The $\bmath{\sigma}\cdot\bmath{L}$ terms in the equations for $F_i$
and $G_i$ are related to pseudospin-orbital and spin-orbital couplings
respectively \cite{arima,meng0,arima2}.

Let us look more closely at the Eq.~(\ref{psiminus}) for $\Psi_-$. The
denominator $E-\Sigma$ in the $\bmath{\sigma}\cdot\bmath{L}$ term
comes from the replacement of $\Psi_+$, using Eq.~(\ref{dirac_first1}), in the term $-[\bmath{\alpha}\cdot\bmath p\,\Sigma]\Psi_{+}$ of Eq.~(\ref{pss}), since
\begin{equation}
\label{psi_plus}
\Psi_+=\frac{\bmath{\alpha}\cdot\bmath p\,\Psi_-}
{E-\Sigma}=
\frac{{\rm i}\,\bmath{\alpha}\cdot\hat r}{E-\Sigma}\bigg(-\frac{\partial\hfill}{\partial r}+
\frac{1}{r} \bmath{\sigma}\cdot\bmath{L}\bigg)\Psi_-\ .
\end{equation}

From Eq.~(\ref{dirac_first1}), when $E-\Sigma=0$, then 
$\bmath{\alpha}\cdot\bmath p\,\Psi_-=0$, meaning that, in Eq.~(\ref{psi_plus}),
the whole term in parenthesis applied to $\Psi_-$ gives zero. Therefore, although each of the two terms within the parenthesis, divided by $E-\Sigma$, is infinite when $E=\Sigma$, their sum is finite. The analysis of these terms is better done by rewriting Eq.~(\ref{pss}) as a Schroedinger-like equation:
\begin{equation}
\frac{p^2}{2m^*}\Psi_{-}+\frac{[\bmath{\alpha}\cdot\bmath p\,\Sigma]}{2 m^*}\Psi_{+}+
\Sigma\,\Psi_-=E\,\Psi_- \ ,
\label{lower_schr}
\end{equation}
where an energy- and $r$-dependent effective mass $m^*=(E+2m-\Delta)/2$ was defined. Applying again Eq.~(\ref{psi_plus}) we have
\begin{equation}
\frac{p^2}{2m^*}\Psi_{-}+\frac{1}{2 m^*}\frac{\Sigma'}{E-\Sigma}\bigg(-\frac{\partial\hfill}{\partial r}+
\frac{1}{r} \bmath{\sigma}\cdot\bmath{L}\bigg)\Psi_-+
\Sigma\,\Psi_-=E\,\Psi_- \ .
\label{lower_schr2}
\end{equation}
Now we can identify the various terms contributing to the binding energy $E$
and, in particular, assess the role of the term $\Sigma'/[2 m^*(E-\Sigma)] \bmath{\sigma}\cdot\bmath{L}/r\,\Psi_-$ in the pseudospin energy splitting. This is possible since the principal value of the integral of this term, after multiplying it
by $\Psi_-^\dagger$, is finite. Furthermore, the
following sum rule should be satisfied:
\begin{equation}
-{\rm P\ }\int\Psi_-^\dagger\frac{1}{2 m^*}\frac{\Sigma'}{E-\Sigma}
\frac{\partial\Psi_-}{\partial r}\,{\rm d}^3{\bmath r}+
{\rm P\ }\int\Psi_-^\dagger\frac{1}{2 m^*}\frac{\Sigma'}{E-\Sigma}
\frac{1}{r} \bmath{\sigma}\cdot\bmath{L}\Psi_-{\rm d}^3{\bmath r}
=\int\Psi_-^\dagger\frac{-{\rm i}\,\bmath{\alpha}\cdot\hat r\,\Sigma'}{2 m^*}
\Psi_+{\rm d}^3{\bmath r}\ ,
\label{sum_rule}
\end{equation}
where {\textquoteleft P\textquoteright} denotes the principal value of the integral.
The analysis of the contribution of the various terms in Eq.~(\ref{lower_schr2}) will be done in
the next section.

It is interesting to discuss how the pseudospin
symmetry gets broken. The SU(2) generators of the pseudospin symmetry
are given by  \cite{levi,smith,bell}
\begin{eqnarray}
S_i&=&s_i\,\frac{1}{2}(1-\beta)+\frac{\bmath\alpha\cdot\bmath{p}\ s_i
\,\bmath\alpha\cdot\bmath{p}}{p^2}\,\frac{1}{2}(1+\beta)=\nonumber\\
&=&\left(\begin{array}{cc}\tilde s_i& 0\\ 0&s_i\end{array}\right)
\end{eqnarray}
where
\begin{equation}
\tilde s_i=\frac{\bmath\sigma\cdot\bmath p}{p} s_i
\frac{\bmath\sigma\cdot\bmath p}{p}=\frac{2\bmath s\cdot\bmath p}{p^2}
p_i-s_i,
\end{equation}
and $s_i={\sigma_i/2}$ ($i=1,2,3$).
 The commutator of this operator with the Hamiltonian (\ref{dirac}) is
\begin{equation}
[H,S_i]=\left(\begin{array}{cc}[\Sigma,\tilde s_i]&0\\ 0&0\end{array}\right)\, .
\label {comut}
\end{equation}
Thus, the pseudospin symmetry breaking can be related to the
commutator $[\Sigma,\tilde s_i]$.  If we take the non-relativistic
limit, we recover the findings of Ref.~\cite{bahri} as long as $\tilde
s_{i}$ commutes with the non-relativistic Hamiltonian $H_{nr}$.  This is
equivalent to consider the helicity transformed spin operator,
$\tilde s_{i}$, to be the generator of a SU(2) symmetry of the
non-relativistic Hamiltonian, $[H_{nr},\tilde s_i]\,=\,0$, implying $[\tilde
H_{nr}, s_i]\,=\,0$, where
\begin{equation}
 \,\tilde H_{nr}\,\,=\,\, \frac{\bmath\sigma\cdot\bmath p}
{p}\,\, H_{nr}\,\, \frac{\bmath\sigma\cdot\bmath p}{p}
\end{equation}
is the helicity transformed non-relativistic Hamiltonian of
Ref.~\cite{bahri}.  Requiring $[\Sigma,\tilde s_i]\,=\,0$ is equivalent to
the previous conditions $\Sigma=S+V=0$ or $d \Sigma/dr=0$, when $\Sigma$
is a radial function \cite{gino,arima,meng0}.

\section{Pseudospin energy splittings for Woods-Saxon mean fields}
 
As stated above, the equivalent conditions $\Sigma=S+V=0$ and 
$d \Sigma/dr=0$ are not met in nuclei. From Ginocchio's 
findings \cite {gino}, one expects that there is a correlation
between the pseudospin splitting and the $\Sigma$ depth.
Our purpose here is to go further and investigate whether there is
also a correlation between those splittings and the shape of
the mean-field nuclear potential, namely its radius and 
surface diffuseness. 

To this end, we perform a model calculation using a Lorentz
structured relativistic potential of Woods-Saxon type in the Dirac
equation.  The scalar and vector components of this potential are the
mean-field central nuclear potentials.  The potentials are given by
\begin{equation}
U(r)=\frac{U_0}{{1+\exp [(r-R)/a]}}\ ,
\label{WSaxon}
\end{equation}
where $U(r)$ stands either for the vector or for the scalar potential.
Although this potential is not a full self-consistent relativistic potential
derived from meson fields, it is realistic enough to
be applied to nuclei. Indeed, most self-consistent potentials
have a Woods-Saxon-like shape, i.e., one can recognize in them a depth, $U_0$, a
radius (range), $R$, and a diffusivity, $a$. Therefore, the study of
pseudospin partners splittings as a function of these parameters
is meaningful and realistic enough to be applied to most nuclei, at least qualitatively. 
It is known that, in certain isotope chains, the
central depth and the surface diffuseness changes sensibly
\cite{nikolaus,meng}. At the same time, as the mass number $A$ increases, the nuclear
radius increases according to the $A^{1/3}$ law,
which means that it is also important to study the role of the parameter $R$ 
in pseudospin symmetry.

Using Woods-Saxon potentials for $\Sigma$ and $\Delta$, we solved numerically
the coupled first-order Dirac equations 
for the radial fields $F_i(r)$ and $G_i(r)$ obtained by factoring out the
angular wave functions of $\Psi_{\pm}$ in Eqs. (\ref{dirac_first}). They read
\begin{subequations}
\label{dirac_first_radial}
\begin{eqnarray}
F'_{\tilde\ell} +\frac{1 + \kappa_{\tilde\ell}}{r} F_{\tilde\ell} & = & 
-(E - m - \Sigma) G_\ell\\
G'_\ell + \frac {1+\kappa_\ell}{r}G_\ell & = & (E + m -\Delta) F_{\tilde\ell}
\end{eqnarray}
\end{subequations}
where
\begin{eqnarray}
\kappa_\ell&=&\left\{\begin{array}{cl}
- (\ell+1) &\quad j =  \ell + \frac{1}{2}\\
\ell &\quad j  =  \ell - \frac{1}{2} \end{array}\right.\\
\kappa_{\tilde\ell} &=& -\kappa_\ell=\left\{\begin{array}{cl}
- (\tilde\ell +1) &\quad  j  = \tilde\ell + \frac{1}{2}\\
\tilde\ell &\quad j  =  \tilde\ell - \frac{1}{2}\end{array}\right. \ .
\end{eqnarray}

There are altogether six parameters for 
$\Sigma$ and $\Delta$, namely the central depths, $\Sigma_0$
and $\Delta_0$, two radii and two diffuseness parameters. 
We observed that the pseudospin splitting is not sensitive to
$R$ and $a$ of the $\Delta$ potential, and, accordingly, set
the same radius, $R$, and surface diffuseness, $a$, for both potentials. 
We first fit these
parameters to the neutron spectra of $^{208}$Pb. The quality
of our fitting is showed in Fig.~1, where the results of the present 
calculation (WS) are shown, together with the experimental values
and those
obtained using  a relativistic mean field approach (model G1 of
Ref.~\cite{furn1}) for the same set of pseudospin doublets:
  $(1i_{11/2}\ ,\ 2g_{9/2})$, $(2f_{5/2}\ ,\ 3p_{3/2})$ and $(1h_{9/2}\ ,\
 2f_{7/2})$, which are the 3 topmost pseudospin partner levels in
 $^{208}$Pb.
\begin{figure}[htb]
\begin{center}
\includegraphics[clip=on,width=7.5cm,angle=-90]{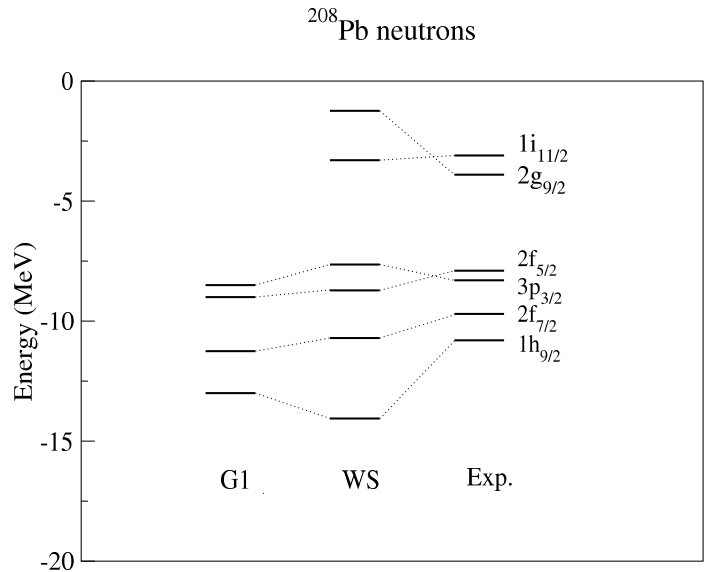}
\end{center}
\caption[Figure 1]{Calculated neutron single-particle energy levels of
the pseudospin partners $(1h_{9/2},\ 2f_{7/2})$,
$(2f_{5/2},\ 3p_{3/2})$ $(1i_{11/2},\ 2g_{9/2})$ in $^{208}$Pb (WS). The
leftmost values are the corresponding values of the model G1 of
Furnsthal {\textit et al.}~\cite{furn1} and the experimental values
are in the rightmost column. The parameters for the Woods-Saxon
potential used to fit the $^{208}$Pb neutron energy levels are
$R=7$ fm, $\Delta_0=650$ MeV, $\Sigma_0=-66$ MeV and $a=0.6$ fm.
The experimental values were taken from Ref.~\cite{campi}.}
\end{figure}

The parameters used for the fit are given in the caption of Fig.~1.
For the present purpose, our fitting is quite reasonable.
Keeping
$\Sigma_{0}$, $\Delta_{0}$, and $R$ fixed, we vary $a$ in order to see
how the energies of the pseudospin doublets are sensitive to the
surface diffuseness.  This dependence is shown in Fig.~2. As $a$
increases, the splittings of the pseudospin doublets decrease.  
This could be expected because the derivative of $\Sigma$ decreases
when $a$ increases. However, by
further increasing $\,a\,$, the pseudospin doublet partners cross
each other, inverting the sign of the energy splitting.  This inversion
of pseudospin partner splittings, $E_{n-1,\tilde l +
1/2} < \,\,E_{n,\tilde l - 1/2}$ changing to $E_{n-1,\tilde l + 1/2} >
\,\,E_{n,\tilde l - 1/2}$, is observed experimentally and was also found
in~\cite{meng0,ring,marcos}. It also occurs for calcium isotopes, analyzed
below.

It is important to note that, once the pseudospin
doublet partners cross each other, i.e., the pseudospin splitting
changes sign, the effect of increasing $a$ is to drive the pseudospin
doublets further apart.
This systematics is consistent with the non-relativistic
prediction of Ref.  \cite{bahri} as shown by the asymptotic
behavior of the radial wave function (\ref{blokhin}).
\begin{figure}[htb]
\begin{center}
\includegraphics[clip=on,width=7.5cm,angle=-90]{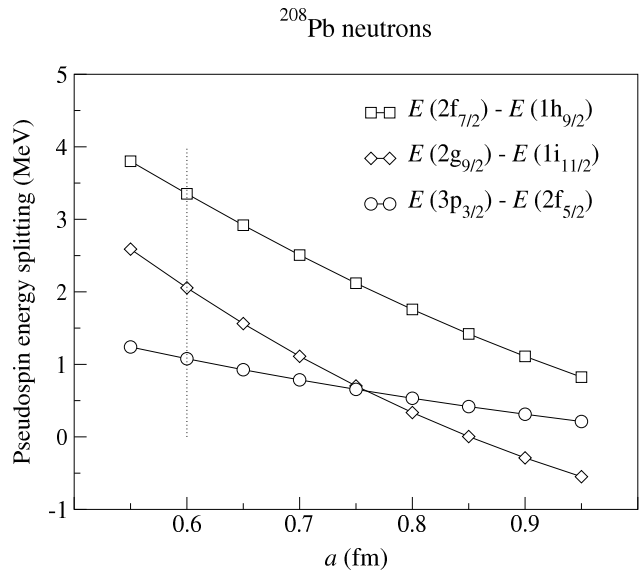}
\end{center}
\caption[Figure 2]{pseudospin energy splittings in $^{208}$Pb as a function of
the diffusivity for the neutron pseudospin partners of Fig.~1. 
The vertical line corresponds to the fitted Woods-Saxon
parameters for $^{208}$Pb given in the caption of Fig.~1.}
\end{figure}

\begin{figure}[htb]
\begin{center}
\includegraphics[clip=on,width=7.5cm,angle=-90]{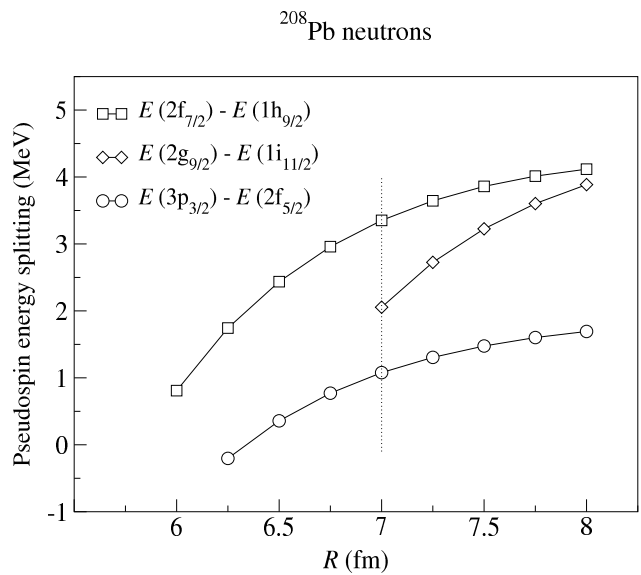}
\end{center}
\caption[Figure 3]{pseudospin energy splittings in $^{208}$Pb as a function of
the radius for the same neutron pseudospin partners of Fig.~1. The vertical line
corresponds to the fitted Woods-Saxon
parameters for $^{208}$Pb given in the caption of Fig.~1.}
\end{figure}

A similar trend is seen when we let the radius $R$ vary, fixing all other 
parameters. The result is shown in Fig.~3. When the radius increase,
the energy splittings increase, which again can be understood by
the dependence of $d\Sigma/ dr$ with $R$. However, we found that the dependence
of the pseudospin splitting on $R$ is
the opposite for the deepest energy levels, which means there are also some
surface effects in this trend. This does not happen when the diffusivity  
changes, the trend being the same for all levels.
Here we can see again the 
phenomenon of level crossing for the pseudospin partners
$(2f_{5/2}\ ,\ 3p_{3/2})$. For $R<7$ fm we could not get bound solutions
for one or both $1i_{11/2}$, $2g_{9/2}$ levels. This dependence on $R$ is
especially important when comparing different isotopes, since, in many cases,
it is the radius of the mean-field potentials that changes most noticeably. 
Finally, we keep $a$, $R$ and $\Delta_0$ fixed but vary $\Sigma_0$ in order
to study the sensitiveness of the pseudospin doublets with the depth of
the central $\,\Sigma\,$ mean field potential.  The results are
presented in Fig.~4 and the following behavior is observed:  as
$|\,\Sigma_0\,|$ decreases, the splitting also decreases.  This is in
accordance with Ginocchio predictions for pseudospin symmetry breaking
due to the finiteness of the $\Sigma$ mean field.
 
\begin{figure}[hbt]
\begin{center}
\includegraphics[clip=on,width=7.5cm,angle=-90]{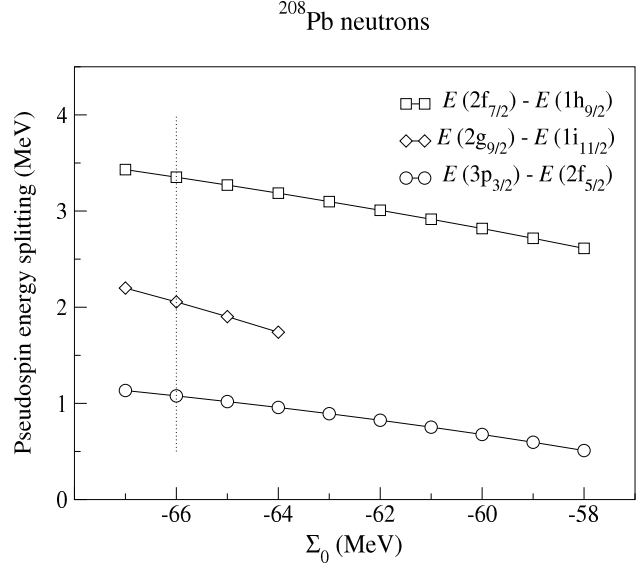}
\end{center}
\caption[Figure 4]{Pseudospin energy splitting in $^{208}$Pb as a
function of the depth of the $\Sigma$ potential for the same
neutron pseudospin partners (similar to Fig.~1). The vertical
line corresponds to the fitted Woods-Saxon parameters for
$^{208}$Pb given in the caption of Fig.~1.}
\end{figure}

From Fig.~4 we see that the decreasing of $|\Sigma_0|$ shifts the levels
$(1i_{11/2},\ 2g_{9/2})$ to the continuum. We also found that,
for a deeper pseudospin doublet, an inversion of pseudospin energy splitting occurs
for sufficiently low $|\Sigma_0|$.

\begin{figure}[ht]
\begin{center}
\includegraphics[clip=on,width=7.5cm,angle=-90]{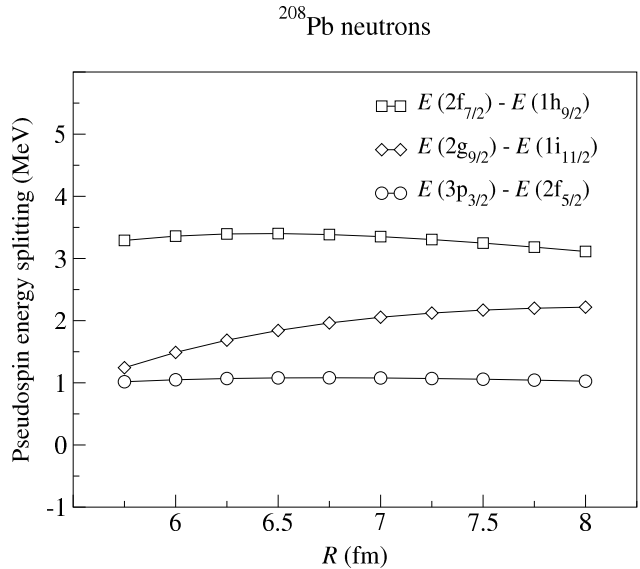}
\end{center}
\caption[Figure 5]{Pseudospin energy splitting in $^{208}$Pb as a
function of $R$ keeping 
$a$ and the product $|\Sigma_0|R^{2}$ fixed for 
 the same neutron pseudospin partners as in Fig.~1.}
\end{figure}

Actually, we have found that $|\Sigma_0|$ and $R$ are not completely
independent parameters.
Varying $|\Sigma_0|$ and $R$ but 
keeping $a$ and the product $|\Sigma_0|R^{2}$ fixed, 
the pseudospin splittings remain almost constant, 
as shown in Fig.~5.

We observed that varying the other free parameter, $\Delta_0$,
does not qualitatively change the splittings.  Since the $\Delta$
potential is related to the effective mass of the nucleons, the basic
effect is to slightly change the nucleon separation energies, and
especially the spin-orbital coupling, as discussed below.

Although the Woods-Saxon potential is expected to be a good
approximation for central potentials of heavy nuclei like lead, some
calculations indicate that it might also be good enough for
lighter nuclei such as calcium \cite{nikolaus}.  Therefore, in order to
show the importance of the surface diffuseness and of the radius in the dynamical
pseudospin symmetry, we applied our model also to two calcium isotopes,
where the pseudospin splitting is also small.
\begin{table}[hbt]
\begin{center}
\begin{tabular}{|c||c|c|c|c|c|c|c|}
\hline
Nucleus & $R$ & $a$ & $\Sigma_0$ & $\Delta_0$& Level & $E_{\rm calc}$ & $E_{\rm exp}$\\
\hline\hline
\rule[-5mm]{0pt}{12mm}$^{40}$Ca & 3.8 & 0.8 & $-$78 & 770 &
$\begin{array}{c} 2s_{1/2}\\1d_{3/2}\end{array}$ & 
$\begin{array}{c}-16.0\\ -15.1\end{array}$&
$\begin{array}{c}-18.2\\ -15.6\end{array}$ \\
\hline
\rule[-5mm]{0pt}{12mm}$^{48}$Ca & 4.0 & 0.8 & $-$64 & 650 &
$\begin{array}{c}2s_{1/2}\\1d_{3/2}\end{array}$ &
$\begin{array}{c}-12.6\\-12.2\\\end{array}$& $\begin{array}{c}-12.4\\ -12.4\end{array}$ \\
\hline
\end{tabular}
\caption[Table 1]{Calculated and experimental energy values of the neutron pseudospin partners $2s_{1/2}$ and
$1d_{3/2}$ in $^{40}$Ca and $^{48}$Ca.
The Woods-Saxon potential parameters used to fit the energy
levels of these nuclei are also displayed. The radius, $R$, and the diffuseness, 
$a$, are in fm and the energies in MeV. The experimental values are from
Ref.~\cite{campi}.}
\end{center}
\end{table}
We fitted our free parameters to the experimental values of the
energies of the topmost neutron levels of $^{40}$Ca and $^{48}$Ca.
The results for the pseudospin partners $( 2s_{1/2}, 1d_{3/2} )$ are
presented in Table 1 and compared with the experimental values.  

The
difference between the $^{48}$Ca neutron potential with respect to the
$^{40}$Ca neutron potential lies basically in 
a smaller modulus of the central depth and in a larger radius.  
We also see that the
magnitude of pseudospin energy splitting decreases from 0.9~MeV
to 0.4~MeV as the number of nucleons $A$ increases.
Since the levels $2s_{1/2}$ and $1d_{3/2}$ are inverted in $^{40}$Ca, 
the effect of the change in $R$ predicted by the systematics is
to increase the difference $E_{2s_{1/2}}-E_{1d_{3/2}}$, or, since this value is
negative, to decrease its magnitude, favoring the symmetry.

We are going now to examine more closely the effect of the term 
$\Sigma'/[(E-\Sigma)(2 m^*)] \bmath{\sigma}\cdot\bmath{L}/r\,\Psi_-$ in 
Eq.~(\ref{lower_schr2}) on pseudospin energy splittings. If we multiply
that equation by $\Psi_-^\dagger$, integrate, and then divide the result by
$\int\Psi_-^\dagger\Psi_-{\rm d}^3{\bmath r}$, we get
\begin{equation}
\bigg\langle \frac{p^2}{2m^*}\bigg\rangle+\big\langle V_{\rm PSO}\big\rangle +
\big\langle V_{\rm D}\big\rangle+\langle\Sigma\rangle=E \ ,
\label{aveg_schroed}
\end{equation}
where
\begin{eqnarray}
\label{kinetic}
\bigg\langle \frac{p^2}{2m^*}\bigg\rangle&=&\frac{\displaystyle \int^\infty_0 dr\,\frac{1}{\displaystyle 2m^*}
\bigg[F_{\tilde\ell}\,\frac{d\hfill}{d r}\bigg(r^2\,\frac { d F_{\tilde\ell}}
{d r}\bigg)+
\tilde\ell(\tilde\ell+1)F_{\tilde\ell}^2\bigg]}
{\displaystyle\int_0^\infty r^2\,dr\,F_{\tilde\ell}^2}\\
\label{pso}
\big\langle V_{\rm PSO}\big\rangle&=&
\frac{\displaystyle{\rm P}\, \int^\infty_0 r\,dr\,
\,\frac{1}{2 m^*}\frac{\Sigma'}{E-\Sigma}(\kappa_\ell-1)F_{\tilde\ell}^2}{\displaystyle\int_0^\infty r^2\,dr\,F_{\tilde\ell}^2}\\
\label{darwin}
\big\langle V_{\rm D}\big\rangle&=&-\,
\frac{\displaystyle{\rm P}\, \int^\infty_0 r^2\,dr\,
\,\frac{1}{2 m^*}\frac{\Sigma'}{E-\Sigma}F_{\tilde\ell}\,
\frac {d F_{\tilde\ell}}{d r}}{\displaystyle\int_0^\infty r^2\,dr\,F_{\tilde\ell}^2}\\
\label{sigma_aver}
\langle\Sigma\rangle&=&\frac{\displaystyle \int^\infty_0 r^2\,\,dr\,\Sigma
\,F_{\tilde\ell}^2}
{\displaystyle\int_0^\infty r^2\,dr\, F_{\tilde\ell}^2\,}\ .
\end{eqnarray}
These terms can be identified, respectively, as a kinetic term, a pseudospin-orbit term,
 a potential term related to what is sometimes called Darwin term
and the mean value of the $\Sigma$ potential with respect to the lower component, $\Psi_-$.
 We computed numerically the principal values, checking that the sum
rule (\ref{sum_rule}) was satisfied.

\begin{figure}[hbt]
\begin{center}
\includegraphics[clip=on,width=7.5cm,angle=-90]{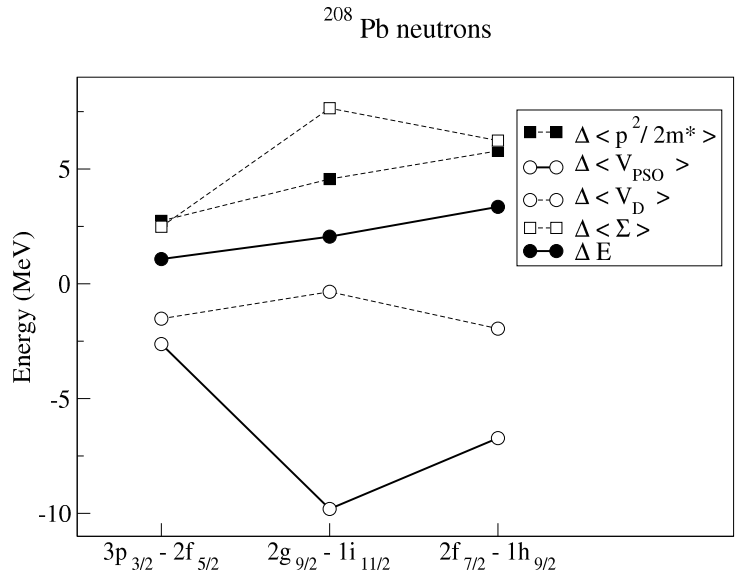}
\end{center}
\caption[Figure 6]{Contribution of the terms (\ref{kinetic}) -- (\ref{sigma_aver})
to the energy splittings $\Delta E$ of the pseudospin partners of Fig.~1. The strong cancelation among the terms produces a small $\Delta E$.}
\end{figure}

The results are shown in Figs.~6, 7 and 8. In Fig.~6 is plotted the difference of the terms in Eq.~(\ref{aveg_schroed}) between 
each member of the pseudospin partners we have been considering. One sees that the contribution for the pseudospin energy splittings 
of the $V_{\rm PSO}$ term is larger than the splitting itself and it is at least of the same order of magnitude and has the
opposite sign of the kinetic and $\langle\Sigma\rangle$ terms. The relative
contribution of each term depends on the particular pseudospin doublet we are considering. This can be seen clearly in figures 7 and 8, 
where the differences of the terms in (\ref{aveg_schroed}) for the pseudospin partners $(1i_{11/2},\ 2g_{9/2})$ and $(2f_{5/2},\ 3p_{3/2})$, 
respectively, are plotted. For states with higher angular momentum
the magnitude of $\big\langle V_{\rm PSO}\big\rangle$ contribution for the energy splitting is higher, since it is roughly proportional to $2\tilde\ell +1$, but there is a significant cancelation with the kinetic
 and  $\langle\Sigma\rangle$ terms. A similar finding was already reported in~\cite{marcos}, where it was pointed out that pseudospin-orbit coupling
cannot be treated as a perturbative quantity. One the other hand, these
figures show a clear correlation between the pseudospin-orbital term and the pseudospin energy splitting when the diffusivity and 
the depth of $\Sigma$ potential are varied.
This feature does not depend on the particular pseudospin doublet under consideration.

\begin{figure}[hbt]
\begin{center}
\includegraphics[clip=on,width=7.5cm,angle=-90]{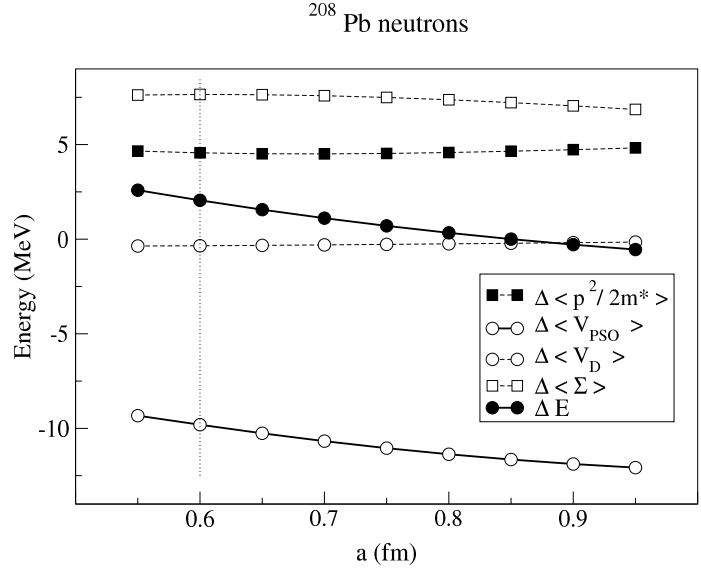}
\end{center}
\caption[Figure 7]{Contribution of the terms (\ref{kinetic}) -- 
(\ref{sigma_aver}) to the energy splitting $\Delta E$ of the pseudospin partner 
$(1i_{11/2},\ 2g_{9/2})$ in $^{208}$Pb as a function of
the diffuseness. The vertical line corresponds to the fitted Woods-Saxon
parameters for $^{208}$Pb given in the caption of Fig.~1.}
\end{figure}

\begin{figure}[hbt]
\begin{center}
\includegraphics[clip=on,width=7.5cm,angle=-90]{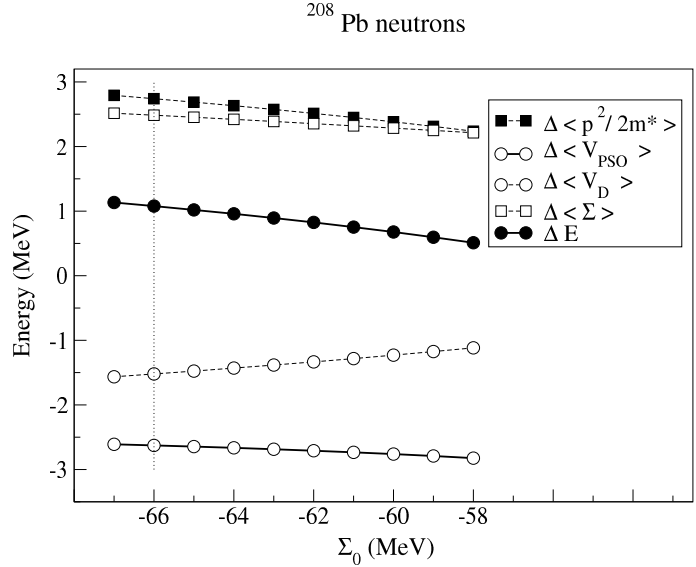}
\end{center}
\caption[Figure 8]{Contribution of the terms (\ref{kinetic}) -- (\ref{sigma_aver}) to the energy splitting $\Delta E$ of the 
pseudospin partner 
$(2f_{5/2},\ 3p_{3/2})$ in $^{208}$Pb as a function of
$\Sigma_0$. The vertical line corresponds to the fitted Woods-Saxon
parameters for $^{208}$Pb given in the caption of Fig.~1.}
\end{figure}

The pseudospin-orbital term ($V_{\rm PSO}$) comes from the
coupling between the upper and the lower components. This term will vanish only 
when the upper component is zero (assuming a non-zero lower component) or 
when $\Sigma'=0$, showing again
that exact pseudospin symmetry is possible only when there are no
bound states. This is not the case for the usual spin-orbit interaction
generated by the term in the square brackets in Eq.~(\ref{ps1}). Here,
this term will vanish when the lower component is zero, which 
corresponds to the non-relativistic limit. 

From the results presented so far, we can conclude that the degeneracy of
pseudospin levels is very much dependent on the shape of the nuclear mean field
potential: the actual choice of the potential parameters,
fitted to describe a nucleus, determines whether or not there is pseudospin symmetry.
Furthermore, it is the strong cancelation presented in Figs.~6--8 that produces the 
quasi-degeneracy of pseudospin levels.
The conclusion is that pseudospin symmetry is a dynamical symmetry in
nuclei, in the sense of Arima's definition of dynamical
symmetry~\cite{arima1}:  (i) a symmetry of the Hamiltonian which is not
geometrical in nature; or (ii) a ordered breaking symmetry from
dynamical reasons. This dynamical nature of pseudospin symmetry is further
stressed by the fact that pseudospin levels can cross each other. We observe
also that the realization of pseudospin symmetry depends on the doublet in 
consideration, although the systematics is the same for all pseudospin partners.

\section{Isospin asymmetry of pseudospin splittings} 

We apply now this systematics to some
nuclei studied in the literature, looking at the differences between
neutron and proton spectra as far as pseudospin symmetry is concerned.

Recently, the pseudospin symmetry in
Zr and Sn isotopes was investigated as a function of the number of
nucleons \cite{meng0}.  The form of the $\Sigma$ potential as a function 
of the radial distance for such nuclei, as $A$ increases, is given in
Ref.~\cite{meng}, starting with $^{100}$Sn and going up to $^{170}$Sn.
This information on $\Sigma$ for Sn allows us to analyze the behavior for
neutron pseudospin doublets going from $A=100$ to $A=170$.  As $A$
increases, the modulus of the central depth, $|\Sigma_0|$, decreases and the
surface diffuseness increases, both effects favoring the pseudospin
symmetry, as shown in Figs.~2 and 3. However, as expected, the radius
$R$ increases with $A$, which can (at least partially) offset the previous
changes.  This is what is observed in Ref.~\cite{meng0}: the energy splittings 
of deep neutron pseudospin partners
decrease but those near the surface do not change sensibly as $A$ 
increases.  The splitting for the partner $(1h_{9/2},\ 2f_{7/2})$
increases from a small negative value to a positive value, i.e., the
levels cross each other.  In \cite{meng0} it is also observed that
pseudospin splittings are smaller for levels close to the nucleus
surface, even crossing each other.  This phenomenon can
also be observed in $^{208}$Pb, as one can see from the experimental
energy values of the pseudospin partners $(2f_{5/2},\ 3p_{3/2})$
$(1i_{11/2},\ 2g_{9/2})$ in Fig. 1.  We do not expect our ansatz
for the nuclear central potentials to explain the behavior of levels
very near the nucleus surface.  Such fine details are outside the
scope of our work, which simply intends to establish a broad
systematics for pseudospin splittings in nuclei.

For Zr isotopes, studied in refs.  \cite{arima,meng0}, although the
authors did not provide details on the potentials, again one sees that
the pseudospin splitting in general decreases with $A$, with a somewhat
irregular behavior for the partner $(1g_{7/2},\ 2d_{5/2})$.  It is
interesting to remark that the splitting for the doublet
($2d_{3/2},3s_{1/2}$) decreases to the point of level crossing.
Therefore, we can conclude that, as $A$ increases, there is a general trend
for the pseudospin doublet partners to approach each other, towards
realization of the symmetry.

The proton spectra of Sn isotopes was also studied in Ref.
\cite{meng}. We observe that the central neutron depth, $\Sigma_0$,
varies from about $-65$ MeV to
$-54$ MeV and the surface diffuseness increases whereas the proton
central depth, $\Sigma_0$, varies from $-50$ MeV to $-60$ MeV and the
surface diffuseness slightly increases. The reason lies on
the $\rho$ meson interaction, which for a neutron rich nuclei is repulsive for the neutrons and
attractive for the protons, as explained by mean field model
calculations \cite{wal}.  The inclusion of this interaction, which is
important in asymmetric nuclei, changes the vector part of the potential
$V$ to
\begin{equation}
\label{V_rho}
V = V_{\omega}+V_\rho=V_{\omega}\pm \frac{g_{\rho}}{2} \rho_0\, ,
\end{equation}
with + and $-$ signs for protons and neutrons respectively; $\rho_0$ is
the time component of the $\rho$ field, which is proportional to the
number of protons minus the number of neutrons, and $V_{\omega}$ comes
from the vector-isoscalar $\omega$ meson. The increase of the parameters
$R$ and $a$ with $A$ can be traced back to the known nuclear
radius $A^{1/3}$ dependence and to the excess of neutrons
on the surface (neutron skin effect). Hence, we may conclude that,
for a given nucleus, the parameters $\Sigma_0$, $a$ and $R$ for
protons and neutrons are different. Then, an isospin asymmetry
in the pseudospin interaction is expected to take place in agreement with the 
analysis of the mean field nuclear parameters for protons and neutrons done
in~\cite{mosk}. 
In particular, since $\rho_0$ is negative for heavy nuclei, the vector
potential
(\ref{V_rho}) is bigger (and, thus, $|\Sigma|$ is smaller) for neutrons
than for protons. From our previous analysis, for a neutron rich nucleus, 
pseudospin symmetry for the neutron spectrum is favored, in agreement
with the results presented by Lalazissis {\em et al.}~\cite{ring}.

The systematics discussed here seems to be quite general and a comment
on how
it affects the spin-orbit splittings is pertinent. Spin-orbit splittings
are much less changed by variations of the central depth and surface diffuseness
of the nuclear potential than pseudospin-orbit splittings.
The reason is that to change substantially spin-orbit splittings one
needs a significant relativistic content, i.e., a large lower
component in the Dirac spinor \cite{palberto}.  The spin-orbit splitting is
completely correlated with the nucleon effective mass \cite{furn2}.
From Eq.~(\ref{upper}), we see that it will depend strongly on $\Delta$.
This potential itself carries a quite
large scale when compared with $\Sigma$ (around a factor ten).
Therefore, the $\rho$ meson potential, $V_{\rho}$, generally about ten
percent of $\Sigma$, becomes irrelevant compared to $\Delta$.  This
justifies why the spin-orbit interaction is roughly isospin symmetric,
being almost the same for neutrons and for protons \cite{meng,chiappa}.

On the contrary, the pseudospin splitting needs less relativistic
content, i.e., a smaller lower component in order to change. This may
explain why the non-relativistic analysis of Ref.~\cite{bahri} of
the origin of the pseudospin symmetry, and manifested by
Eq.~(\ref{blokhin}), works well to explain the small pseudospin
splitting.  The pseudospin orbit interaction depends on $\Sigma$, as
one can see from Eq.~(\ref{lower}). This means that now $V_{\rho}$ 
cannot be neglected, since its values are comparable to those of $\Sigma$,
causing sensibly different variations of the central depth for neutrons
and protons, for which $V_{\rho}$ has different signs.  This is one of
the mechanisms generating the isospin asymmetry in the pseudospin-orbit
interaction, as we have pointed out before.
This can be seen in Fig.~9, where the 
energy splittings for the pseudospin partners $(1h_{9/2},\ 2f_{7/2})$,
$(2f_{5/2},\ 3p_{3/2})$ and for spin-orbit pairs $(2f_{5/2},\ 2f_{7/2})$,
$(1h_{9/2},\ 1h_{11/2})$ are plotted. Here we have just fixed the radius and diffusivity,
using the values that fitted the neutron spectrum of $^{208}$Pb,
and let both $\Sigma_0$ and $\Delta_0$ change by the same amount, thereby
simulating the increase of the vector potential $V$ that affects $\Sigma$ and
$\Delta$ in the same way. 
It is clear that the splittings change much less
for the spin-orbit doublets than for the pseudospin doublets.
The magnitude of the change depends on the specific level one considers, but we checked
that this kind of behavior occurs for deep levels. Of course, one should
also consider the effect of the Coulomb potential for proton spectra, which has 
the opposite sign of $V_\rho$ for protons. However, we expect the effect of 
$V_\rho$ to be dominant for neutron rich nuclei. 

\begin{figure}[hbt]
\begin{center}
\includegraphics[clip=on,width=7.5cm,angle=-90]{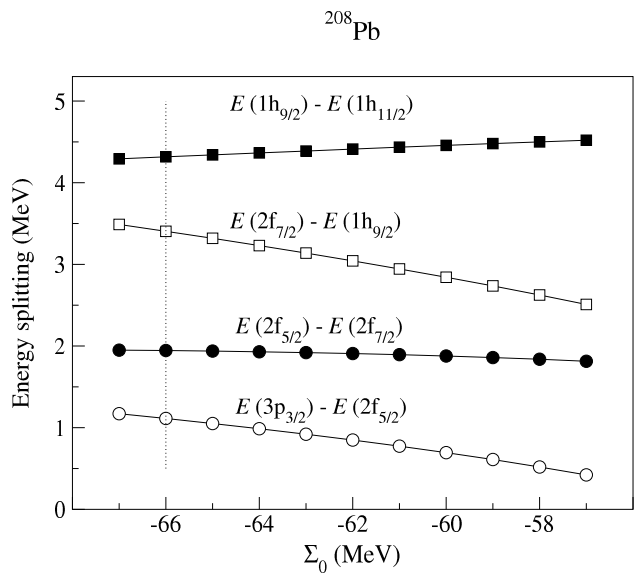}
\end{center}
\caption[Figure 9]{Energy splitting in $^{208}$Pb as a
function of the depth of the $\Sigma$ potential for the 
pseudospin partners $(1h_{9/2},\ 2f_{7/2})$,
$(2f_{5/2},\ 3p_{3/2})$ (hollow symbols) and spin-orbit pairs $(2f_{5/2},\ 2f_{7/2})$,
$(1h_{9/2},\ 1h_{11/2})$ (filled symbols). The depth $\Delta_0$ varies 
from 639 MeV to 649 MeV so that only the vector potential changes. The vertical line corresponds to the fitted Woods-Saxon parameters for
$^{208}$Pb given in the caption of Fig.~1.}
\end{figure}

\section{Conclusions}

We have studied, through a mean-field model calculation
with Woods-Saxon potentials, the role
played by the surface diffuseness, the radius, and the central depth
of the sum of the standard vector and scalar nuclear
potentials in the energy splittings of pseudospin partners, both for
lead and calcium isotopes.
This study allowed us to draw a systematics for the behavior of the
pseudospin splitting as $a$, $R$ and $\Sigma_0$ vary. 
Such behavior was confirmed when we
applied the systematics to the Sn nuclei isotope chain, predicting the
observed general trend towards degeneracy of pseudospin doublets.
Based on our findings and in general features of the neutron and proton
potentials, we explained the observed isospin asymmetry of pseudospin
splittings.  
We were able to identify a pseudospin-orbital term and found that its
contribution to the pseudospin energy splittings is large.
The near degeneracy of pseudospin levels is obtained from a significant cancelation among the different terms
 in the Schroedinger-like equation for the lower component of the Dirac spinor. We also
 showed a clear correlation between the pseudospin-orbital term and the pseudospin
energy splitting when the diffusivity and the depth of $\Sigma$ potential are varied. 
 These findings led us to argue that pseudospin
symmetry in nuclei is dynamic.

We believe that the systematics observed in our model calculation and the non-perturbative nature of 
the pseudospin-orbit term, which is larger than the energy splitting itself and needs to be canceled in 
order to produce the pseudospin quasi-degeneracy may help to
understand how the pseudospin symmetry is dynamically broken in nuclei.

\begin{acknowledgments}
We acknowledge financial support from FCT (POCTI), Portugal, and
from CNPq/ICCTI Brazilian-Portuguese scientific exchange program.
\end{acknowledgments}



\end{document}